\documentclass[reprint,twocolumn]{revtex4}

\usepackage{graphicx}
\usepackage{rotating}
\usepackage{array}
\usepackage{amsmath}
\usepackage{amsfonts}
\usepackage{amssymb}
\usepackage{multirow}
\usepackage{mathtools}
\usepackage{txfonts}
\usepackage{bm}
\usepackage{color}

\DeclareGraphicsExtensions{.pdf}

\begin{document}
	
\title{Grid-based diffusion Monte Carlo for fermions without the fixed-node approximation}
\author{Alexander A. Kunitsa} \email{aakunitsa@gmail.com}
\author{So Hirata} 
\affiliation{Department of Chemistry, University of Illinois at Urbana-Champaign, Urbana, Illinois, 61801, USA}
	
\begin{abstract}
A diffusion Monte Carlo algorithm is introduced that can determine the correct nodal structure of the wave function of a few-fermion system
and its ground-state energy without an uncontrolled bias. This is achieved by confining signed random walkers to the points of a uniform infinite spatial grid, allowing them to meet and annihilate one another to establish the nodal structure without the fixed-node approximation. An imaginary-time propagator is derived rigorously from a discretized Hamiltonian, governing a non-Gaussian, sign-flipping, branching, and mutually annihilating random walk of particles. The accuracy of the resulting stochastic representations of a fermion wave function is limited only by the grid and imaginary-time resolutions and can be improved in a controlled manner. The method is tested for a series of model problems including fermions in a harmonic trap as well as the He atom in its singlet or triplet ground state. For the latter case, the energies approach from above with increasing grid resolution and converge within $0.015~{E}_\text{h}$ of the exact basis-set-limit value with a statistical uncertainty of $10^{-5}~{E}_\text{h}$ without an importance sampling or Jastrow factor. 
\end{abstract}
\maketitle

\section{Introduction}

Stochastic algorithms \cite{kalos_monte_1962,anderson_randomwalk_1975,luchow_quantum_2011} hold exceptional promise in treating correlated electronic structures owing to their high parallel efficiency, 
near-exact accuracy, scalability with the problem size, and tiny memory footprints.  Perhaps, the most successful of such algorithms is diffusion Monte Carlo (DMC) \cite{anderson_randomwalk_1975}, but it is notoriously 
plagued by an uncontrolled bias (the fixed-node error) arising from the fixed-node approximation \cite{klein_nodal_1976,reynolds_fixednode_1982,anderson_randomwalk_1975} introduced 
as a practical solution to the sign problem \cite{troyer_computational_2005}. It amounts to using the nodal structure of some trial wave function, 
which differs from the exact one, thus causing the error.

The objective of this work is to eliminate the fixed-node error from DMC \cite{anderson_randomwalk_1975}.
This is achieved by confining the positively and negatively signed walkers on an infinite, uniform, real-space grid, which can thus meet 
on a grid point and then annihilate one another, establishing a nodal structure without the fixed-node or any other similar approximation. The resulting nodal structure 
should converge at the exact one in the limit of infinitesimally small grid spacing and imaginary time step.
A general stochastic propagation protocol on a grid is derived in this work. 
Our method---grid DMC---is distinguished from any of the previously developed fermion quantum Monte Carlo approaches that enforce annihilation 
by walker pairing, correlated dynamics, or other techniques \cite{carlson_mirror_1985,coker_dmc_1986,james_anderson_quantum_1991,kalos_correlated_1996,kalos_model_1997,kalos_exact_2000,mishchenko_remedy_2006}. 

Casula {\it et al.}\ \cite{michele_casula_diffusion_2005,casula_size-consistent_2010} were among 
 the first to introduce a real-space grid in DMC, but for the different purpose of implementing a nonlocal pseudopotential. Like their method, 
our grid DMC relies on a finite-difference approximation to the kinetic-energy operator on a grid. 
 It also shares some algorithmic features with full-configuration-interaction quantum Monte Carlo (FCIQMC), which propagates signed walkers in a discretized 
 space of the Slater determinants, the latter ensuring the fermion antisymmetry of the wave function.
Grid DMC obeys a similar population dynamics as FCIQMC, which is an interplay between propagation, branching, and annihilation of walkers \cite{spenser_sign_2012}. 
A correct nodal structure emerges as the total number of walkers exceeds a critical value, $N_\text{c}$, which is a function of the grid spacing, 
dimension of the configuration space, and character of the target state. The value of $N_\text{c}$ determines the memory footprint, 
which is expected to grow exponentially with the system size as a manifestation of the sign problem \cite{troyer_computational_2005}.

Clearly, grid DMC is severely limited in its applicability because of the exponential size-dependence of $N_\text{c}$, 
if one insists on
solving the Schr\"{o}dinger equation essentially exactly.
In this work, we establish its feasibility just for two-particle (Coulomb) systems in the 3-D space with a view to ultimately realizing DMC-like stochastic algorithms
for two-electron theories such as second-order M{\o}ller--Plesset perturbation (MP2) \cite{sinanoglu_1962,bischoff_computing_2013,hirata_numerical_2017}
or coupled-cluster doubles (CCD) theory \cite{kottmann_coupled-cluster_2017}. It may be argued that stochastic MP2 and CCD 
are expected to be more scalable than their deterministic counterparts (with respect to both the number of processors and problem size) 
and thus applicable to large systems that do not lend themselves to local-correlation speedup.

This article is structured as follows: A detailed description of our grid-DMC algorithm is given in Sec.\ \ref{sec:theory_grid}. 
The results of demonstrative calculations on fermions in a harmonic trap and the He atom in the triplet or singlet ground state are reported in Secs.\ \ref{subsec:bench1} and \ref{subsec:bench2}, respectively. A summary of the method and an outline of the future work are given in Sec.\ \ref{sec:summary}.
In Appendices \ref{sec:pd_derivation} and \ref{sec:proofs}, mathematical details of the algorithm are given.  

\section{Theory and Algorithms}
\label{sec:theory_grid}
Let $\Psi(\bm{r}_1, \dots, \bm{r}_N,\tau)$ be the wave function of an $N$-particle system satisfying the imaginary-time-dependent 
Schr\"{o}dinger equation with energy offset $\omega$,
\begin{eqnarray}
\label{eqn:SE}
-\frac{\partial \Psi(\bm{r}_1, \dots, \bm{r}_N,\tau)}{\partial \tau} = \left(\hat{H} - \omega \right)\Psi(\bm{r}_1, \dots, \bm{r}_N,\tau),
\end{eqnarray}
where $\hat{H}$ is the Hamiltonian with a local, spin-independent potential $V(\bm{r}_1, \dots, \bm{r}_N)$,
\begin{eqnarray}
\label{eqn:H}
	\hat{H} = -\frac{1}{2} \sum_{i = 1}^N \nabla_i^2 + V(\bm{r}_1, \dots, \bm{r}_N).
\end{eqnarray}

The transition probability of the particles hopping from $(\bm{r}_1, \dots, \bm{r}_N)$ to $(\bm{r}_1^\prime, \dots, \bm{r}_N^\prime)$ in short time $\tau$ is approximated by 
the following Green's function, using the Suzuki--Trotter expansion \cite{trotter_product_1959}:
\begin{eqnarray}
\label{eqn:G}
&& G(\bm{r}_1, \dots, \bm{r}_N \to\bm{r}_1^\prime, \dots, \bm{r}_N^\prime ; \tau) 
\nonumber\\ &&\approx
\langle \bm{r}_1^\prime, \dots, \bm{r}_N^\prime | \exp\{-\tau(V - \omega)/{2}\} | \bm{r}_1^\prime, \dots, \bm{r}_N^\prime \rangle 
\nonumber\\&& \times\,
\langle \bm{r}_1^\prime, \dots, \bm{r}_N^\prime | \sum_{i=1}^N \exp(\tau \nabla_i^2/2) | \bm{r}_1, \dots, \bm{r}_N \rangle
\nonumber\\&& \times\,
\langle \bm{r}_1, \dots, \bm{r}_N | \exp\{-\tau(V - \omega)/2\} | \bm{r}_1, \dots, \bm{r}_N \rangle \\
&& = \exp\left\{-\tau \left( \frac{ V(\{\bm{r}^\prime_i\}) + V(\{\bm{r}_i\}) }{2} - \omega \right)\right\}
\nonumber\\&& \times\, \prod_{i=1}^N \langle \bm{r}_i^\prime|  \exp(\tau \nabla_i^2/2) | \bm{r}_i \rangle,
\label{eqn:G2}
\end{eqnarray}
where $|\bm{r}_1,\dots,\bm{r}_N\rangle$ and $|\bm{r}_i\rangle$ are the position eigenfunctions.

In this work, a particle is confined to a point in an infinite, uniform, 3-D grid with grid spacing $\delta$. 
The Laplacian in the kinetic-energy operator is approximated by a central three-point finite-difference formula.
In this ansatz, each factor in the kinetic-energy part of the Green's function, Eq.\ (\ref{eqn:G2}), 
simplifies to
\begin{eqnarray}
&& \langle \bm{r}^\prime|  \exp(\tau \nabla^2/2) | \bm{r} \rangle 
\nonumber\\&& \approx
\langle x^\prime, y^\prime, z^\prime |\exp(\tau \nabla^2/2)| x, y, z \rangle \\
&& =
\langle x+n_x \delta, y+n_y \delta, z+n_z \delta |\exp(\tau \nabla^2/2)| x, y, z \rangle \\
&& = p_{n_x} p_{n_y} p_{n_z}
\end{eqnarray}
with 
\begin{eqnarray}
p_{n} &=& \left \langle x+n\delta  \left|\exp\left( \frac{\tau}{2}\frac{\partial^2}{\partial x^2}\right)\right| x \right\rangle \\
&=& \frac{1}{2\pi} \int_{-\pi}^{\pi} \cos(kn)\exp\left\{-\frac{2\tau}{\delta^2}\sin^2\left(\frac{k}{2}\right)\right\} dk,
\label{eqn:ke_prop_fin}
\end{eqnarray}
where it should be understood that $(x, y, z)$ and $(x^\prime, y^\prime, z^\prime)$ are grid points and $n_x$, $n_y$, and $n_z$ are integer displacements.
For the derivation of Eq.\ (\ref{eqn:ke_prop_fin}), see Appendix \ref{sec:pd_derivation}.

Evidently, $p_n$ is the non-Gaussian transition probability of a particle hopping from a grid point to its $n$th nearest neighbor (where $n$ 
is a positive or negative integer) in an infinite, evenly-spaced, 1-D grid. It satisfies the following properties expected of such probability:
\begin{eqnarray}
p_n &=& \delta_{n0}\,\,\text{for }\tau=0, \label{eqn:cond1}\\
p_n &\geq& 0, \label{eqn:cond2}
\end{eqnarray}
and
\begin{eqnarray}
\sum_{n=-\infty}^{\infty} p_n &=& 1. \label{eqn:cond3}
\end{eqnarray}
The proofs of these identities can be found in Appendix \ref{sec:proofs}. 

\begin{figure}[!tbh]
	\begin{center}
		\includegraphics[width=6cm]{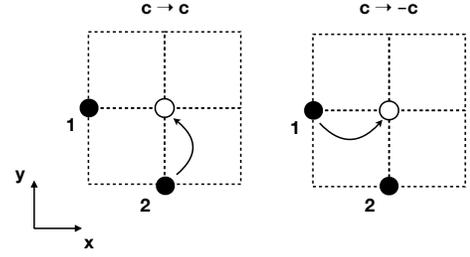}
		\caption{Examples of sign-preserving (left) and sign-flipping (right) moves for two fermions (labeled `1' and `2') with the same spin 
		on a 2-D grid. The grid points are canonical ordered in the increasing $x$ coordinates first, and then in the increasing $y$ coordinates. In the move shown on the left, the coordinates of particles 1 and 2 are in a canonical order before and after the move, preserving the sign of $c$. The move shown on the right brings the coordinates of particles 1 and 2 into a non-canonical order, which needs to be permuted once to be canonical-ordered,
		causing a sign flip in $c$.}
		\protect\label{fig:sign}
	\end{center}
\end{figure}

Equation (\ref{eqn:G}) contains moves essentially corresponding to an interchange of particles with the same spin, which should, therefore, reverse its sign when applicable. 
This can be encoded by introducing a canonical order of the grid points and associating the Green's function with the parity of the permutation that brings 
the sequence of the destination grid points into a canonical order. In this work, we define a canonical order as one with the increasing $x$ coordinates first, then
with the increasing $y$ coordinates, and finally with the increasing $z$ coordinates. Examples of sign-preserving and sign-flipping moves are depicted 
in Fig.\ \ref{fig:sign} for a simple case of two same-spin fermions on a 2-D grid. 

The foregoing equations admit a stochastic implementation similar to DMC \cite{anderson_randomwalk_1975,luchow_quantum_2011} with a major difference being in the definition of random walkers and their (signed) transition probabilities. 
Each walker represents a set of all particles, carrying a unit signed weight, $c = \pm 1$. The $i$th particle has spatial coordinates that are on a point in an infinite, uniform grid with a grid spacing of $\delta$ and spin label $s_i = \pm 1/2$ (in the case of an electron) so that $\sum_{i=1}^N s_i = S_z$, where $S_z$ is the total magnetic spin angular momentum quantum number of the target state.
The walker weight $c$ flips sign, whenever an odd number of permutations is needed to bring the particle coordinates into a canonical order after a move. 
The calculation workflow is similar to that of DMC and consists of the following steps:
\begin{enumerate}
\item \textbf{Initialization.} An initial walker population is randomly generated from some real trial wave function $|\Psi_\text{T}(\bm{r}_1,\dots, \bm{r}_N)|$
(which must not be confused with a trial wave function in the fixed-node approximation) by the Metropolis algorithm \cite{Metropolis}, exercising care to avoid Coulomb singularities.  
The initial walker weights are assigned with the sign of $\Psi_\text{T}$. Factors $p_n$ of Eq.\ (\ref{eqn:ke_prop_fin}) are computed as a function of $n$ by numerical integration 
and stored. 

\item \textbf{Propagation.} Each walker performs a random walk. 
This step is executed by looping over all particles $(i = 1, \dots, N)$ and displacing the grid coordinates of each particle relatively by $(n_x, n_y, n_z)$ with a transition probability of $p_{n_{x}} p_{n_{y}} p_{n_{z}}$. The walker weight $c = \pm 1$ is then multiplied by $(-1)^{[P]}$, where $[P]$ is the parity of permutation $P$ that brings the sequence of the new particle coordinates into a canonical order. 

\item \textbf{Branching.} For each walker, the old $(\bm{r}_1,\dots,\bm{r}_N)$ and new  $(\bm{r}^\prime_1,\dots,\bm{r}^\prime_N)$ 
sets of the particle coordinates are used to calculate the branching factor, $m = \exp[-\tau\{(V(\{\bm{r}^\prime_i\}) + V(\{\bm{r}\}))/2 - \omega\}]$. The walker is subsequently replaced by $\lfloor m + \xi \rfloor$ copies, where $\xi$ is a random number sampled from a uniform distribution on the interval $[0,1]$.  

\item \textbf{Annihilation.} The list of all walkers is sorted and then searched for the members whose particles occupy the identical set of grid points. 
An equal number of positively and negatively signed walkers among them are annihilated, leaving only a minimal number of like-signed walkers on each set 
of grid points.

\item \textbf{Energy estimation.} The energy offset $\omega$ is updated as $\omega \rightarrow \omega + \tau^{-1} \ln (N_\text{w}/N_\text{w}^\prime)$ to keep the number of walkers approximately constant, where $N_\text{w}$ and $N_\text{w}^{\prime}$ are the sizes of the old and new walker lists, respectively. The mean value of $\omega$ in the limit $\tau \to \infty$ can be used to estimate the ground-state energy,
\begin{eqnarray}
\label{eqn:growth_en}
E_\text{gr}= \langle \omega \rangle,
\end{eqnarray}   
which is known as the growth estimator \cite{umrigar_diffusion_1993}. A more desirable measure of energy according to statistical uncertainty considerations is the projection estimator \cite{an_fixed-node_1991,booth_fermion_2009} defined as follows:
\begin{eqnarray}
\label{eqn:proj_en}
E_\text{proj} = \left \langle \frac{\sum_k c_k \hat{H} \Psi_\text{T} (\bm{r}^{[k]}_1, \dots, \bm{r}^{[k]}_N) }{\sum_k c_k \Psi_\text{T} (\bm{r}^{[k]}_1, \dots, \bm{r}^{[k]}_N) } \right \rangle
\end{eqnarray}   
at $\tau \to \infty$, where $c_k$ and ($\bm{r}^{[k]}_1,\dots,\bm{r}^{[k]}_N$) are the weight and grid points of the $k$th walker, 
and $\Psi_\text{T}$ is any trial wave function (which may differ from the one used in Step 1) having a non-zero overlap with the exact wave function. 
\end{enumerate} 

Steps 2 through 5 are repeated (with each cycle counted as one Monte Carlo step) until convergence. 
For a sufficiently large number of walkers, $N_\text{w}$, the computational cost of a Monte Carlo cycle is 
dominated by the annihilation step, which exhibits $N_\text{w} \ln N_\text{w}$ scaling of computational
complexity owing to the need to sort the walker list. 
As will be shown in the next section, another important implication of the annihilation is the existence of the critical number of walkers, $N_\text{c}$, 
required to obtain the correct nodal structure, which grows exponentially with the dimension of the configuration space. 
The value of $N_\text{c}$ defines the memory footprint and limits the application size. 
The lack of annihilation for a small number of walkers, i.e., $N_\text{w} < N_\text{c}$, leads to node sampling errors and introduces a nodal bias in the  energy estimates \cite{booth_fermion_2009}. 

\section{Results and Discussion}

A Python program of grid DMC was written and made available online \cite{kunitsa_gridDMC}.
The transition probabilities $p_n$ were calculated (with a maximum error of $10^{-10}$) using the Clenshaw--Curtis adaptive quadrature implemented in the {\sc gnu} scientific library \cite{gsl24}. 
Displacements with the probabilities smaller than $10^{-8}$ were discarded. 
Statistical errors in the energies were evaluated using the blocking analysis \cite{flyvbjerg_error_1989} as implemented in {\sc pyblock} module \cite{spencer_pyblock}. 

\subsection{Fermions in a harmonic trap}
\label{subsec:bench1}
Consider a system of four noninteracting spin-1/2 particles with a unit mass confined in a 1-D harmonic trap characterized by a potential, $V =  \sum_{i=1}^4 x_i^{2}/2$, where $x_i$ is the $i$th particle coordinate. It lends itself to analytical solution of the Schr\"odinger equation with the exact energy, $E = \sum_{i=1}^4 (n_i + 1/2)$ in the units of 
$E_\text{h}$, where $n_i$ is the $i$th quantum number of the 1-D harmonic oscillator which is a nonnegative integer.
States with the total spin quantum number $S=0$, $1$, and $2$ were studied by grid DMC with a grid spacing of $\delta=0.1$ a.u.\ and an imaginary time step of $\tau=0.1$ a.u.
The initial particle positions were sampled from a uniform distribution on a 6.0-a.u.\ interval centered at the origin. An ensemble of $\sim10^{7}$ random walkers was propagated over 5000 Monte Carlo steps. The energy was evaluated by the growth estimator. Table \ref{table:spin_states_4} compiles the results.

The $S=0$ ground state has a pair of $\alpha$- and $\beta$-spin particles occupying the $n=0$ one-particle harmonic-oscillator level and another $\alpha\beta$ pair occupying the $n=1$ level, having the exact energy of $4.0~E_\text{h}$.   The $S=1$ ground state has two particles in the $n=0$ level and one $\alpha$-spin particle each in
the $n=1$ and $2$ levels, whereas the $S=2$ ground state has one $\alpha$-spin particle each in the $n=0$, $1$, $2$, and $3$ levels. 
The correct nodal structure of the $n\geq 1$ one-particle wave functions and the overall antisymmetry of the four-particle wave function
naturally emerge in these grid-DMC calculations. 
Unlike DMC with the fixed-node approximation, where the wave function in each nodal packet has the correct shape, but not throughout the whole space
even after adjusting the sign of each packet, the wave function obtained in grid DMC has the correct shape in the whole space.

The grid-DMC energies are accurate within $0.006$, $0.01$, and $0.02~E_\text{h}$ of the exact values of the $S=0$, $1$, and $2$ 
states, respectively. They are many orders of magnitude greater than the statistical uncertainty of $\sim 5\times 10^{-5}~E_\text{h}$, and so they are biases.
The main sources of the biases are the nonzero grid spacing and finite time step. When the deterministic calculations were performed using the same grid 
(i.e., the diagonalization of the discretized Hamiltonian on a uniform grid spanning 6 a.u.\ using the central three-point finite-difference formula with $\delta = 0.1$\ a.u.), 
their energies are within $0.002 \sim 0.004~E_\text{h}$ of the grid-DMC results. These remaining biases are attributed to the finite time step.

\begin{table}[h]
	\caption{Growth-estimator energies (statistical uncertainties in parentheses) in $E_\text{h}$ of the $S=0$, $1$, and $2$ states of the system with four noninteracting spin-1/2 fermions with a mass of 1 a.u.\ in a 1-D harmonic trap.}
	\label{table:spin_states_4}
		\begin{tabular}{cccc}
			\hline\hline
			$S$ &  Grid DMC\tablenotemark[1] & Grid\tablenotemark[2] & Exact\tablenotemark[3] \\
			\hline
			0  & $3.99458(4)$ & 3.99625 & 4.0 \\
			1  & $4.99168(4)$ & 4.99374  & 5.0 \\
			2  & $7.98292(5)$ & 7.98622  & 8.0 \\
			\hline \hline
		\end{tabular}
		\tablenotetext[1]{A grid spacing of $0.1$\ a.u.\ and an imaginary time step of $0.1$\ a.u.}
		\tablenotetext[2]{Central three-point finite-difference method with a grid spacing of $0.1$\ a.u.}
		\tablenotetext[3]{Analytical results.}
\end{table}

\begin{figure}[!tbh]
	\begin{center}
		\includegraphics[width=8cm]{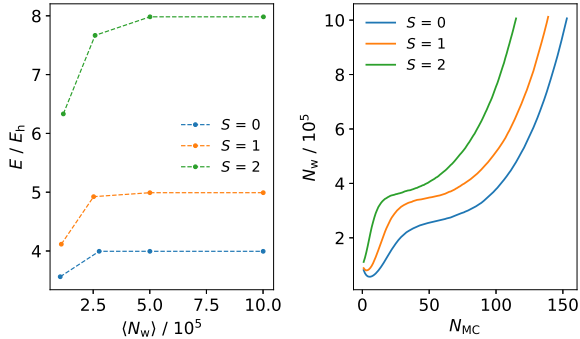}
		\caption{Left:\ the energy evaluated by the growth estimator, $E$, as a function of the average number of walkers, $\langle N_\text{w}\rangle$, for the $S=0$, $1$, and $2$ ground states of the four noninteracting fermions in a 1-D harmonic trap. Right:\ The number of walkers, $N_\text{w}$, as 
		a function of the Monte Carlo steps, $N_\text{MC}$. The energy onset, $\omega$, was held fixed at 4.25, 5.25, and 8.25~$E_\text{h}$ for the states with $S=0$, $1$, and $2$, respectively.} 
		\protect\label{fig:con_walkers}
	\end{center}
\end{figure}

In order to study the efficacy of the walker annihilation, we performed a series of runs with varying walker ensemble sizes. The simulation results are 
given in Fig.\ \ref{fig:con_walkers}. The left panel shows that the energies converge at the correct limits, provided that the annihilation events are sufficiently frequent or, equivalently, the number of walkers exceeds a threshold value $N_\text{c}$, which varies with the character of the target state. The lack of a sufficient number of annihilation events causes the underestimation
of the energies. The rate of convergence tends to decrease for higher spin states, in which particles occupy one-particle levels with more nodes. 
The population dynamics presented in the right panel, obtained by holding $\omega$ fixed (a production run adjusts $\omega$), 
is reminiscent of that obtained in FCIQMC \cite{booth_fermion_2009}, similarly exhibiting pronounced plateaus 
signaling that $N_\text{w}$ has reached $N_\text{c}$.  The plateaus occur because of the competition between the spawning and walker annihilation \cite{spenser_sign_2012}.
 
\begin{figure}[!tbh]
	\begin{center}
		\includegraphics[width=8cm]{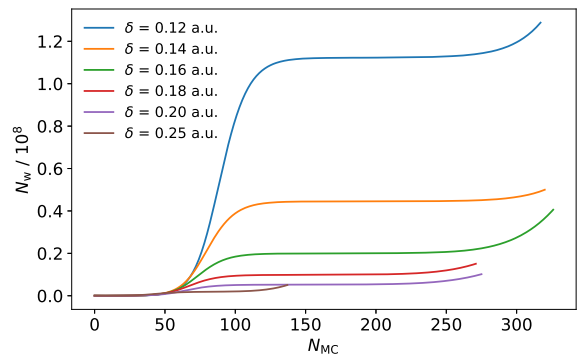}
		\caption{The number of walkers, $N_\text{w}$, as a function of the Monte Carlo steps, $N_\text{MC}$, in grid DMC for two spin-1/2 fermions in a 3-D harmonic trap in the triplet ground state 
		for several grid spacings $\delta$. 
		The energy onset was held fixed at $\omega = 4.5~E_\text{h}$ and imaginary time step $\tau = 0.1$ a.u.}
		\protect\label{fig:pop_2e_gr}
	\end{center}
\end{figure}

The dependence of $N_\text{c}$ on the grid spacing was further analyzed for two spin-1/2 fermions in a 3-D harmonic trap in the triplet ground state (Fig.~\ref{fig:pop_2e_gr}). 
Owing to the higher dimension of the configuration space, $N_\text{c}$ is orders of magnitude larger than in the 1-D system and exhibits a steep growth with decreasing 
$\delta$. A least-squares fit suggests that  $N_\text{c} \approx 339 \,\delta^{-5.99}$. In general, one could expect $N_\text{c} \propto \delta^{-nd}$, where $n$ is the number of particles and $d$ is the dimension of the configuration space. 

\subsection{The He atom in the $^3S$ and $^1S$ states}
\label{subsec:bench2}

The grid-DMC algorithm for a system with charged particles requires some modifications to avoid the Coulomb singularities. 
In contrast to continuous DMC, random walks in a discretized space have a nonzero probability of 
exact particle coalescence, leading to divergent branching factors. To avoid this, we placed the He nucleus at coordinates $(\delta/2, \delta/2, \delta/2)$, 
where $\delta$ is the grid spacing, and furthermore barred the random walks that result in electron-electron coalescence. 
Various grid spacings $\delta$ in the range of 0.01 to 0.16 a.u.\ were tested, while an imaginary time step of 0.005 a.u.\ was used in all calculations.

For comparison, we also performed grid-DMC calculations for the $^3S$ state with the fixed-node approximation using the exact nodal structure \cite{dario_bressanini_unexpected_2005} as well as for the nodeless $^1S$ ground state \cite{Ceperley1991}. The nodal constraint was imposed by killing walkers attempting to acquire a sign inconsistent with that of a trial wave function $\Psi_\text{T}$ having the exact nodal structure. Namely, the walker weight $c_k$ was set to zero if $c_k \Psi_\text{T}  (\bm{r}_1^{[k]}, \bm{r}_2^{[k]}) < 0$. With the assistance of the exact nodal structure, grid-DMC calculations with the fixed-node approximation 
need only 10000 walkers to converge. 

The first six rows of Table \ref{table:he_triplet} list the results of grid-DMC calculations with and without the exact nodal constraint using 
two grid spacings:\ $\delta = 0.16$ or $0.08$ a.u. In both cases, the convergence with respect to the number of walkers was ensured by repeating the calculation with different walker ensemble sizes and checking the stability of the energy estimates. Additionally, the average fraction ($w$) of walkers with the correct sign (i.e., with the same sign as 
$\Psi_\text{T}$) was recorded to quantify the accuracy of the nodal structure. The closer the values of $w$ to100\%, 
the more accurate the nodal structure of the grid-DMC result without the fixed-node approximation.

With a coarse grid of $\delta = 0.16$ a.u., convergence is achieved with $3.5$ to $7\times10^7$ walkers (the number identified as $N_\text{c}$)
with $97.9$ to $98.3$\% accurate nodal structure. The corresponding energies ($-2.1269~E_\text{h}$) have minuscule statistical uncertainties of $10^{-5}~E_\text{h}$, but 
suffer from a much greater bias of $1~\text{m}E_\text{h}$ from the one with the exact nodal constraint ($-2.1278~E_\text{h}$). 
This bias is clearly due to an inexact nodal structure and may be called a nodal bias. 
The grid-DMC result with the exact nodal constraint is too high as compared with the exact energy 
($-2.1753~E_\text{h}$) \cite{pekeris_he_1959} by $48~\text{m}E_\text{h}$, which must be 
ascribed to a combination of nonzero grid spacing and finite time step. As will be shown below, a majority of this bias 
is due to the grid spacing and may be called a grid bias. 

Halving the grid spacing to $\delta = 0.08$ a.u.\ increases $N_\text{c}$ by an order of magnitude
to $6 \times 10^8$. The proportion of walkers with the correct sign deteriorates slightly to 96.7 to 96.8\% 
instead of improves. Interestingly, however, the nodal bias in the energies seems to be compressed to $0.5~\text{m}E_\text{h}$, although 
this value is obscured by the statistical uncertainty of a similar size in the grid-DMC calculation with the exact nodal constraint. 
Nevertheless, it may be concluded that grid DMC can achieve 97-98\% accurate nodal structure {\it a priori}
with a submillihartree nodal bias in the energy for two fermions in the 3-D space. Before an extrapolation to $\delta =0$ limit (see below),
it achieves the energy ($-2.1613$ to $-2.1617~E_\text{h}$) of the He atom in the $^3S$ state within $15~\text{m}E_\text{h}$ of the exact nonrelativisitc value \cite{pekeris_he_1959} without the Jastrow factor or importance sampling.

Further halving the grid spacing to $\delta = 0.04$ a.u.\ might 
elevate the estimated value of $N_\text{c}$ to $3.8\times10^{10}$, approaching a hardware memory limit.

\begin{table}[h]
	\caption{Projection-estimator energies, $E$ (statistical uncertainties in parentheses) 
	of the He atom in the $^3S$ and $^1S$ states obtained by grid DMC (with an imaginary time step of $0.005$\ a.u.)\ with and without an exact nodal constraint.
	$\delta$ is the grid spacing, $\langle N_\text{w} \rangle$ is the average number of walkers, and $w$ is the proportion of walkers having the correct sign.}
	\label{table:he_triplet}
		\begin{tabular}{clcccl}
			\hline\hline
			State & Method & $\delta~/~\text{a.u.}$ & $\langle N_\text{w} \rangle$ & $w~/~\%$ & \multicolumn{1}{c}{$E~/~E_\text{h}$}  \\
			\hline
			$^3S$ & Grid DMC & 0.16  & $3.5 \times 10^7$ & $97.9$  & $-2.12695(1)$   \\
			$^3S$ & Grid DMC & 0.16  & $7.0 \times 10^7$  & $98.3$&  $-2.12687(1)$    \\
			$^3S$ & Grid DMC (fn\tablenotemark[1]) & 0.16  & $10^4$   & $100$ &  $-2.1278(8)$ \\
			$^3S$ & Grid DMC & 0.08  & $6.16 \times 10^8$ & $96.7$ & $-2.16126(1)$  \\
			$^3S$ & Grid DMC & 0.08  & $6.25 \times 10^8$  & $96.8$ & $-2.16169(1)$ \\
			$^3S$ & Grid DMC (fn\tablenotemark[1]) & 0.08  & $10^4$  & $100$ &  $-2.1612(15)$  \\
			$^3S$ &Grid DMC (fn\tablenotemark[1]) &  0.04 & $10^4$ & $100$&  $-2.1698(8)$ \\ 
			$^3S$ & Grid DMC (fn\tablenotemark[1]) & 0.02 & $10^4$ & $100$&  $-2.1724(10)$ \\
			$^3S$ & Grid DMC (fn\tablenotemark[1]) & 0.01 & $10^4$ & $100$&  $-2.1739(7)$ \\
			$^3S$ & Grid DMC (fn\tablenotemark[1]) & 0 \tablenotemark[2] & $10^4$   & $100$   & $-2.1741$ \\
			$^3S$ & Exact\tablenotemark[3] & $\dots$ & $\dots$ & $\dots$ & $-2.1753$ \\
$^1S$ & Grid DMC & 0.16 & $10^4$ & $100$&  $-2.8355(22)$ \\ 
$^1S$ & Grid DMC & 0.08 & $10^4$ & $100$&  $-2.8867(14)$ \\
$^1S$ & Grid DMC  & 0.04 & $10^4$ & $100$&  $-2.8984(14)$ \\ 
$^1S$ & Grid DMC & 0.02 & $10^4$ & $100$&  $-2.9032(16)$ \\
$^1S$ & Grid DMC & 0.01 & $10^4$ & $100$&  $-2.9029(15)$ \\
$^1S$ & Grid DMC & 0\tablenotemark[2]      & $10^4$ & $100$&  $-2.9035$ \\
$^1S$ & Exact\tablenotemark[3] & $\dots$ & $\dots$ & $\dots$& $-2.9037$ \\

			\hline \hline
		\end{tabular}
		\tablenotetext[1]{Fixed-node approximation using the exact nodal structure.}
		\tablenotetext[2]{Extrapolation by a least-squares fitting of the fixed-node results to a quadratic function of $\delta$.}
		\tablenotetext[3]{Exact nonrelativistic energies due to Pekeris \cite{pekeris_he_1959}.}
\end{table}

\begin{figure}[h]
	\begin{center}
		\includegraphics[width=8cm]{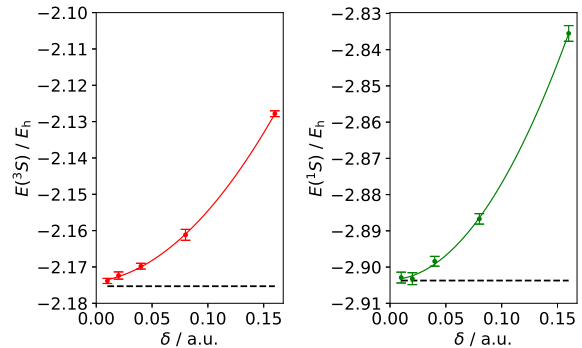}
		\caption{The energies of the He atom in the $^3S$ (left) and $^1S$ (right) states obtained by grid DMC with the exact nodal constraint 
		as a function of the grid spacing ($\delta$). Statistical errors are shown with vertical bars. 
		Horizontal dashed lines indicate the exact nonrelativistic energies \cite{pekeris_he_1959}. 
		Quadratic fit of the grid-DMC energies are drawn with solid curves.} 
		\protect\label{fig:fn_3s_he}
	\end{center}
\end{figure}

The remainder of the calculations in Table \ref{table:he_triplet} were performed with the exact nodal constraint
for the $^3S$ and $^1S$ states with different grid spacings ($\delta$) to estimate a grid bias. Only 10000 walkers were
necessary to converge the energies within a few millihartrees.
They are plotted as a function of $\delta$ 
in Fig.\ \ref{fig:fn_3s_he}. 

For both states, the plots underscore the slow convergence of the energies to the exact nonrelativistic energies \cite{pekeris_he_1959}.
The slow convergence can be attributed to the inherent inefficiency of a uniform grid in describing electron-nuclear and to a lesser extent electron-electron cusps \cite{bischoff_computing_2013}, which may, therefore, be alleviated by the Jastrow or other explicit-correlation factor.
Interestingly, the grid acts to suppress the fluctuations in the branching factor by preventing the walkers from closely approaching the nucleus and give rise to the order of magnitude smaller statistical errors as compared to similar continuous DMC calculations \cite{coker_dmc_1986}. Note that the bias 
appears to decrease monotonically and seemingly quadratically with $\delta$ in the range from 0.16 to 0.01 a.u. 
For finer grids, the character of convergence is hard to discern as it is masked by statistical uncertainties.  
A least-squares fitting of the energies to a quadratic function of $\delta$ can extrapolate the energy of each state at $\delta=0$ that is within $1~\text{m}E_\text{h}$ of 
the respective exact value. This also suggests that the bias is nearly entirely due to a nonzero grid spacing and much less to a finite time step.

\section{Summary and Outlook}
\label{sec:summary}

We developed a novel grid-based quantum Monte Carlo algorithm for a many-fermion wave function with arbitrary nodal structure without invoking the fixed-node approximation. To this end, the original continuous DMC formulation was mapped onto its lattice counterpart by representing the Hamiltonian and corresponding Green's function 
on an infinite uniform spatial grid using a central finite-difference approximation for the kinetic-energy operator. 
We showed that the associated propagator is similar to that of the continuous DMC and reduces to it in the limit of zero grid spacing, yet describing 
a non-Gaussian branching and annihilating random walks of fermions. A key component of the formalism is the definition of a canonical order of particle coordinates, which allows
the algorithm to unambiguously encode the antisymmetry of the many-fermion wave function. 

For a series of systems, we demonstrated that our grid-DMC algorithm managed to converge to the correct nodal structure {\it a priori} 
provided that a total number of walkers exceeded a critical value $N_\text{c}$. 
The latter determines the memory footprint of the method and restricts the applicability to low-dimensional model problems with smooth potentials.
However, we showed that the correct nodal structure and energy of the He atom in the $^3S$ state could be determined {\it a priori} with accuracy of 
97-98\% and 99.4\%, respectively, without an importance sampling or Jastrow factor. The number of walkers needed was $10^7$ to $10^9$, not exceeding
17~GB of memory if 64-bit integers are used to store walker coordinates on a grid. Furthermore, the remaining bias in the energy seems to be nearly entirely caused by the grid spacing and can be effectively removed by
extrapolation \cite{mckoy_numerical_1968}. This opens a possibility of developing practical, scalable, DMC-like stochastic algorithms for two-electron theories
widely used in quantum chemistry and solid state physics such as MP2 and CCD for large systems, which may furthermore include 
an importance sampling and Jastrow factor in their final forms.

\section{Acknowledgments}

The authors thank Professor David M. Ceperley for helpful discussions. 
This work was supported by the U.S.\ Department of Energy, Office of Science, Basic Energy Sciences under Grant No.\ DE-SC0006028. 
It is also part of the Blue Waters
sustained-petascale computing project, which is supported by the National
Science Foundation (awards OCI-0725070 and ACI-1238993) and the state of
Illinois.  Blue Waters is a joint effort of the University of Illinois at
Urbana-Champaign and its National Center for Supercomputing Applications.

\appendix
\section{Derivation of Eq.\ (\ref{eqn:ke_prop_fin})}
\label{sec:pd_derivation}

We approximate the second derivative in the kinetic-energy operator by the central three-point finite difference:
\begin{eqnarray}
    \label{eqn:lapl_delta}
&&-\frac{1}{2} \frac{\partial^2}{\partial x^2} \exp(ikx)  
\nonumber\\&& 
\approx-\frac{1}{2}\frac{\exp\{ik(x + \delta)\} + \exp\{ik(x - \delta)\} - 2 \exp(ikx)}{\delta^2}
\nonumber\\&& 
 = -\frac{1}{2}\frac{\exp(ik\delta) + \exp(-ik\delta) - 2}{\delta^2}  \exp(ikx)
\nonumber\\&& 
 = \frac{2}{\delta^2}\sin^2\left( \frac{k\delta}{2}\right) \exp(ikx),
\end{eqnarray}
where $\delta$ is the grid spacing. 
The transition probability or Green's function for the kinetic-energy operator on a uniform 1-D grid then becomes
\begin{eqnarray}
p_n
&=& \left\langle x+n\delta \left | \exp\left(\frac{\tau}{2} \frac{\partial^2}{\partial x^2}\right)\right |  x \right \rangle
\nonumber\\
&=& \frac{\delta}{2\pi}\int_{-\pi/\delta}^{\pi/\delta}  \exp\{-ik(x+n\delta)\} \exp\left(\frac{\tau}{2} \frac{\partial^2}{\partial x^2}\right) \exp(ikx)\,dk
\nonumber\\
&\approx& \frac{\delta}{2\pi}\int_{-\pi/\delta}^{\pi/\delta}  \exp(-ikn\delta)\exp\left\{-\frac{2\tau}{\delta^2} \sin^2\left(\frac{k\delta}{2}\right)\right\} dk \label{eqn:appendix}
\\
&=& \frac{1}{2\pi}\int_{-\pi}^{\pi}  \exp(-ik^\prime n)\exp\left\{-\frac{2\tau}{\delta^2} \sin^2\left(\frac{k^\prime}{2}\right)\right\} dk^\prime, \label{eqn:appendix2}
\end{eqnarray}
which is identified as Eq.\ (\ref{eqn:ke_prop_fin}). Note that the integration domain $[-\pi/\delta,\pi/\delta]$ is the first Brillouin zone under the periodic boundary
condition with lattice constant $\delta$.

The eigenvalue of the discretized kinetic-energy operator reduces to the correct continuous-space limit as $\delta \to 0$:
\begin{eqnarray}
\lim_{\delta \to 0}\frac{2}{\delta^2}\sin^2\left( \frac{k\delta}{2}\right) = \frac{k^2}{2}.
\end{eqnarray}
Using this and applying the saddle point approximation \cite{mathews_mathematical_1970} to Eq.\ (\ref{eqn:appendix}), we find
\begin{eqnarray}
\label{eqn:gauss_prob}
\lim_{\delta\to0} p_n &=& \frac{\delta}{2\pi} \int_{-\infty}^{\infty} \exp(-ikn\delta) \exp\left(-\frac{\tau k^2}{2} \right) dk  \\
&=& \frac{\delta}{\sqrt{2\pi\tau}}\exp\left( -\frac{n^2\delta^2}{2\tau}\right), \label{eqn:gauss}
\end{eqnarray}
which is a Gaussian function dictating diffusion in a continuous space. Therefore, taking the limit $\delta \to 0$ in grid DMC, 
we recover the usual continuous DMC \cite{anderson_randomwalk_1975,luchow_quantum_2011}.

\begin{figure}[h]
	\begin{center}
		\includegraphics[width=8cm]{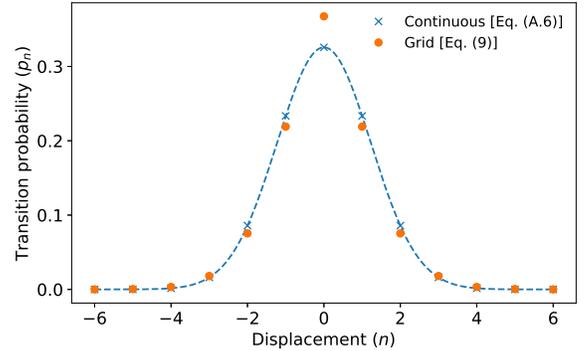}
		\caption{Comparison of $p_n$ [Eq.\ (\ref{eqn:ke_prop_fin}) or (\ref{eqn:appendix2})] versus Gaussian function [Eq.\ (\ref{eqn:gauss})] with 
		$\tau =0.015$ and $\delta = 0.1$.} 
		\protect\label{fig:gau_numer}
	\end{center}
\end{figure} 

Figure \ref{fig:gau_numer} plots the transition probability $p_n$ [Eq.\ (\ref{eqn:ke_prop_fin}) or (\ref{eqn:appendix2})] and its $\delta = 0$ limit (a Gaussian function) [Eq.\ (\ref{eqn:gauss})] as a function of $n$. The transition probability on a grid differs visibly from the Gaussian function in the $\delta = 0$ limit especially for small $n$. 
Specifically, dividing the space into bins slightly increases the probability of staying in the same bin at expense of decreasing the probability to hop to the 
nearest or second nearest neighbors. For a greater displacement, the two plots converge because $\delta$ becomes small relative to the displacement, making 
the saddle point approximation asymptotically exact. At every $n$, $p_n$ is found nonnegative. 

Equation (\ref{eqn:ke_prop_fin}) can, in principle, be generalized for any symmetric finite-difference formula. However, 
for such a higher-order formula, $p_n$ is usually no longer positive for all $n$ unless $\delta^2 \ll {\tau}$ (in which case the space is effectively continuous).  
It is also to be observed that our approach is not immediately extensible to an importance sampling transformation as usually performed in the context of DMC~\cite{reynolds_fixednode_1982} because it breaks the Hermitian and translational symmetry of the kinetic-energy operator 
by introducing a drift term proportional to the first derivative of the wave function.

\section{Proofs of Eqs.\ (\ref{eqn:cond1})--(\ref{eqn:cond3})}
\label{sec:proofs}

A proof of Eq.\ (\ref{eqn:cond1}) is trivial. 

We have
\begin{eqnarray}
p_n &=& \frac{1}{2\pi} \int_{-\pi}^{\pi} \cos(kn)\exp\left\{-\frac{2\tau}{\delta^2}\sin^2\left(\frac{k}{2}\right)\right\} dk 
\nonumber\\ 
&=& \frac{1}{2\pi n} \int_{-\pi n }^{\pi n} \cos(k^\prime)\exp\left\{-\frac{2\tau}{\delta^2}\sin^2\left(\frac{k^\prime}{2 n}\right)\right\} dk^\prime 
\nonumber,
\end{eqnarray}
where $k^\prime = kn$. 
It then follows for $n\to\pm\infty$
\begin{eqnarray}
p_n 
&\approx& \frac{1}{2\pi |n|} \int_{-\infty}^{\infty} \exp(ik^\prime) \exp\left(-\frac{\tau k^{\prime2}}{2\delta^2 n^2}\right) dk^\prime \\
&=& \frac{\delta}{\sqrt{2\pi\tau}}\exp\left( -\frac{n^2\delta^2}{2\tau}\right) \ge 0. \label{eqn:appB1}
\end{eqnarray}
Equation (\ref{eqn:appB1}) imply that $p_n$ and $p_{n+1}$ are nonnegative for a sufficiently large $|n|$.

In the meantime, a recursion relationship for $p_n$ can be derived with integration by parts. For $n \neq 0$,
\begin{eqnarray}
p_n &=& \frac{1}{2\pi} \int_{-\pi}^{\pi} \cos(kn)\exp\left\{-\frac{2\tau}{\delta^2}\sin^2\left(\frac{k}{2}\right)\right\} dk 
\nonumber\\
&=& \frac{1}{2\pi} \left[ \frac{\sin(kn)}{n} \exp\left\{-\frac{2\tau}{\delta^2}\sin^2\left(\frac{k}{2}\right)\right\} \right]^{\pi}_{-\pi}
\nonumber\\
&& +  \frac{1}{2\pi} \int_{-\pi}^{\pi} \frac{\sin(kn)}{n} \frac{\tau \sin(k)}{\delta^2} \exp\left\{-\frac{2\tau}{\delta^2}\sin^2\left(\frac{k}{2}\right)\right\} dk 
\nonumber\\
&=&  \frac{\tau}{2\pi n \delta^2} \int_{-\pi}^{\pi} \sin(kn)\sin(k) \exp\left\{-\frac{2\tau}{\delta^2}\sin^2\left(\frac{k}{2}\right)\right\} dk 
\nonumber\\
&=&  \frac{\tau}{n \delta^2} \frac{p_{n-1} - p_{n+1}}{2},
\end{eqnarray}
which can be rearranged to yield
\begin{eqnarray}
p_{n-1} =  \frac{2 n \delta^2}{\tau} p_{n} + p_{n+1}.
\end{eqnarray}
This proves Eq.\ (\ref{eqn:cond2}) for all $n$ by mathematical induction: Starting from a sufficiently large $n$ that renders both 
$p_n$ and $p_{n+1}$ nonnegative, $n$ is decremented down to zero (vice versa for negative $n$). 
The foregoing also implies that $p_n$ is monotonically decreasing with $n \ge 0$. 

Equation (\ref{eqn:cond3}) can be proven with the Fourier transform of Dirac's $\delta$ function,
\begin{eqnarray}
\sum_{n=-\infty}^{\infty} p_n &=& 
\int_{-\pi}^{\pi} \left\{ \frac{1}{2\pi} \sum_{n=-\infty}^{\infty} \cos(kn) \right\} \exp\left\{-\frac{2\tau}{\delta^2}\sin^2\left(\frac{k}{2}\right)\right\} dk
\nonumber\\ &=& \int_{-\pi}^{\pi} \delta(k) \exp\left\{-\frac{2\tau}{\delta^2}\sin^2\left(\frac{k}{2}\right)\right\} dk = 1.
\end{eqnarray}

 \bibliography{abbr.bib,akunitsa_qmc.bib}
\end{document}